# DISTRIBUTION NETWORK FAULT PREDICTION UTILISING PROTECTION RELAY DISTURBANCE RECORDINGS AND MACHINE LEARNING

Ebrahim BALOUJI
Eneryield – Sweden
balouji@eneryield.com

Karl BÄCKSTRÖM
Eneryield – Sweden
kalle@eneryield.com

Viktor OLSSON
Eneryield – Sweden
viktor@eneryield.com

Petri Hovila
ABB – Finland
petri.hovila@fi.abb.com

Henry NIVERI
ABB – Finland
henry.x.niveri@fi.abb.com

Anna KULMALA
ABB – Finland
anna.kulmala@fi.abb.com

Ari SALO
Vaasan Sähköverkko – Finland
ari.salo@vaasansahkoverkko.fi

## ABSTRACT

*As society becomes increasingly reliant on electricity, the reliability requirements for electricity supply continue to rise. In response, transmission/distribution system operators (T/DSOs) must improve their networks and operational practices to reduce the number of interruptions and enhance their fault localization, isolation, and supply restoration processes to minimize fault duration. This paper proposes a machine learning-based fault prediction method that aims to predict incipient faults, allowing T/DSOs to take action before the fault occurs and prevent customer outages.*

## INTRODUCTION

The increasing scope and complexity of electrical power systems across all sectors, including generation, transmission, distribution, and load systems, is resulting in a higher frequency of faults. The most common types of grid faults are partial or complete short circuits of power lines to the ground or among themselves. These faults can lead to significant financial losses and decrease the reliability of the electrical system. Utilities and large industrial plants, which often have extensive power line systems, are prone to faults due to various reasons, including:

- Aging and wear and tear of power lines during operation
- Using power lines that are not suitable for the intended application
- Mechanical failure, such as damage to power lines during installation or subsequent use
- Degradation of power line sheaths and insulation due to e.g., extreme temperatures, chemicals, weather, or abrasion
- Moisture build-up in insulation
- Electrical overloading
- Birds or other animals
- Vegetation that is too close or trees falling on power lines

Early fault prediction can significantly benefit grid operators by enabling them to address potential issues before they lead to failures. This can improve the overall reliability of the grid, resulting in decreased operational costs and reduced revenue loss, as well as ensuring the continuity of power delivery to end users. Before a fault happens, the grid often suffers from precursor symptoms, making it possible to predict faults using appropriate models that detect these symptoms. There are some existing studies on the prediction and analysis of grid faults, which we will briefly discuss here. One of the most well-known approaches to predict power line failures, that has been studied over the past few years, is by analysing partial discharges [1-5]. However, this analysis requires expensive measurement tools and is very challenging in noisy situations, which is often the case in large-scale grids. Another drawback with this approach is that it is not feasible in the case of fault prediction in underground and underwater cables. Additionally, not all faults originate from insulation degradation, but due to other reasons, like the ones mentioned above, which the partial discharge-based methods are not suitable for.

Other methods for predicting faults have also been investigated in recent years, such as using temperature sensors on power lines [6-9] and monitoring power lines with unmanned aerial vehicles [10-12]. However, these methods have limitations in detecting faults in underground and underwater cables and are also limited in their ability to detect faults originating from some root causes, making it necessary to use multiple approaches in order to cover all potential faults. Furthermore, these existing approaches require the deployment of additional sensors, which can be a costly endeavour as it involves not only the cost of the sensors themselves, but also the establishment of low-latency and high-throughput data stream communication pipelines, processing and storage.

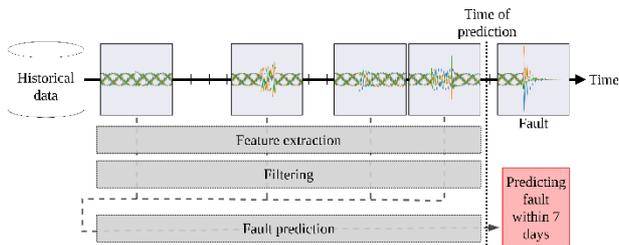

Figure 1. The diagram provides an overview of the forecasting pipeline. The process begins by collecting anomalous measurements of current and voltage. A set of relevant features are extracted from each measurement, and any irrelevant measurements are filtered out. The filtered sequence of anomalies is then passed through a machine learning prediction model, which generates a warning if a fault is imminent.

In this paper, we propose a machine learning-based approach that (i) predicts faults in transmission and distribution power lines in up to a week before they happen, (ii) utilizes a purely data-driven approach by applying feature engineering methods as a pre-processing step before feeding data to a long short-term memory (LSTM) based deep neural network and (iii) utilizes existing measurements of only current and voltage without requiring installation of any additional sensors. A high-level schema of the developed method is shown in Figure 1, the detail of each part will be explained in detail in the next section.

## METHODOLOGY

### Data

The data used for prediction is collected without the need for additional sensors, using the same instrument transformers and I/O connections as the protection and control devices in the example installation. Current and voltage measurements and I/O statuses of primary equipment are shared using IEC 61850-9-2LE and IEC 61850 8-1 protocols with an edge computing device. The edge device runs multiple high-sensitivity protection-related algorithms [13], which trigger the recording of disturbances for use in the prediction.

After the edge device has collected the data from the substation, the prediction is made in a more computationally efficient environment. Disturbance recordings and other relevant data are uploaded to a cloud-based service for monitoring and storage. The data from this central storage is then transferred to a dedicated machine learning service, where the actual prediction is performed. The results of the prediction are then transferred back to the monitoring service for contextual processing. This approach of pre-filtering on the edge device significantly reduces the amount of data transferred to the cloud compared to a setup where all accurate measurements are sent to the cloud, which results in a more efficient use of computational resources.

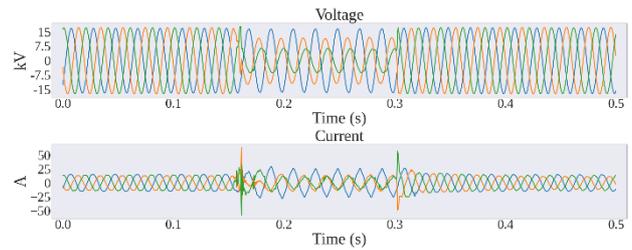

Figure 2. An example of an anomaly recorded by the protection relay. The recording includes a short snippet of three phases of voltage and current waveforms.

The data collection is done in a medium voltage (20kV) substation, and the prediction is made for feeders, including underground cables and overhead lines. The disturbance recordings are recorded voltage and current waveforms with a sampling rate of 4 kHz, see an example in Figure 2.

### Feature Engineering

The waveform recordings are high-dimensional and contain a large amount of information. As a result, feature engineering is necessary before inputting the data into the machine learning model. Each disturbance recording is processed to extract several derived features, such as RMS, impedance, active and reactive power, harmonics, and phase angles. From these signals, several representative scalar values are calculated, such as max, min, standard deviation, in order to include as much information from the disturbance recording as possible with a minimal number of values. This processing results in a set of highly descriptive scalars arranged in a feature vector with approximately 300 values, serving as a lower-dimensional representation of the waveform recording, which can be more easily processed by the machine learning-based prediction model. The feature extraction pipeline is illustrated in Figure 3.

### Machine Learning

The feature vectors are arranged to create a multidimensional time series with dimensions *(# recordings, # feature values)*. These time series are then used as input to the forecasting model. For a given prediction, the input data is taken from a one-week-long period, meaning that all events happening within a time window stretching from the time of prediction and one week back are included in the input. The task is illustrated in figure 4; when making a prediction for a different time, the input time window is moved.

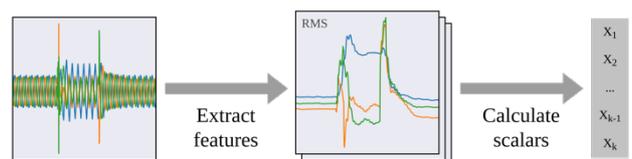

Figure 3: The data pre-processing pipeline. Several relevant features are derived from the recorded waveforms. From these features, various scalar values are calculated and arranged in a feature vector to be used as input to the machine learning model.

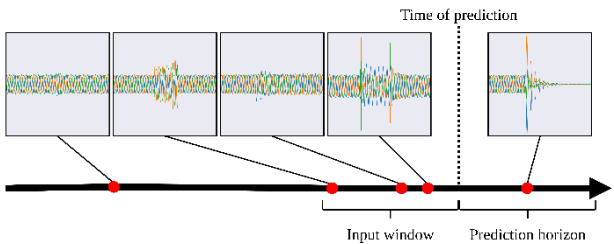

Figure 4: The input to the prediction machine learning model is a sequence of feature vectors, each corresponding to a disturbance recording within a time window before the time of prediction. In this illustration, the data corresponding to the leftmost recording would not be included in the input, while the next three would. The target is a binary signal describing whether or not there will be a fault within a prediction horizon from the time of prediction. In this figure, the target would be one, since a fault occurs shortly after the time of prediction.

The target when training the model is a binary signal that at each time is one if there is a fault occurring within a predefined time period after that time, and zero otherwise. Specifically, this time period is also one week. This means that the model's output, which is a number between zero and one, can be interpreted as the probability of a fault occurring within one week. Thus, the model warns about faults up to one week before they happen.

Our forecasting model is a neural network that consists of three main components: a filtering component, an LSTM layer, and a classification head. The filtering component is a fully connected neural network that classifies each recording in the input time series as relevant or not relevant to the forecasting process. We manually labeled a subset of the recordings as either containing an actual anomaly, or a normal state that can be recorded if the recording device sensitivity is high and which is not very relevant to the forecasting. The filtering component is then trained on these labeled samples to learn how to distinguish between relevant and irrelevant recordings. Only relevant recordings are passed on to the next stage of processing. The LSTM (Long Short-Term Memory) layer is a type of recurrent neural network that is well-suited for handling irregularly sampled time series data due to its ability to selectively remember or forget certain pieces of information from the input sequence [14]. It processes the input time series and generates a single feature vector. The output is then processed through a classification head consisting of two fully connected layers and projected with a sigmoid activation function to a value between 0 and 1, representing the probability of a fault occurring within the next week. The model is visualized in figure 5.

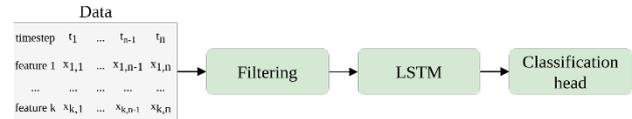

Figure 5: An overview of the architecture of the forecasting machine learning model. The data is arranged in a time series, where each time step corresponds to a data recording. The data is first filtered to remove irrelevant recordings. Then the time series is processed by an LSTM and passed to a classification head, giving a probability of an imminent fault as output.

If data from multiple connected locations is available, the network can utilize that information by separately processing a time series for each location and concatenating the outputs from the LSTM layer before jointly passing them to the classification head. This allows the model to separately process each location, while at the same time selectively pay attention to relevant information in each location.

We apply various augmentations to the data before it is used for training, i.e., add transformed versions of the data to our data set in order to change its variability. One significant augmentation is shuffling the input data such that the three phases change order, e.g., treating phase one as phase two, etc. This effectively increases the number of samples by six times when all permutations are used. This is crucial as it addresses the limitation of having a small amount of data.

## RESULTS AND DISCUSSION

### Evaluation methodology

The results shown in this section are based on if the output of our model is larger or smaller than 0.5 and if the corresponding target value is zero or one, i.e., if there will be a fault within one week. Predictions are made once each hour. To evaluate the model, we split the data into a training set, and a test set, with the latter used to assess the model's ability to generalize to unseen data. We also used 5-fold cross-validation, which involves partitioning the data into five equal-sized sets, training the model on four sets, and evaluating its performance on the remaining set. This process was repeated for each of the five folds, and the average performance across all iterations was calculated as the final result.

### Hyperparameter tuning

We present a limited hyperparameter tuning experiment to investigate the effect of learning rate and learning rate decay on the performance of the forecasting model. The learning rate is a crucial parameter that controls the speed at which the weights of the neural network are updated during training, while learning rate decay is a technique that gradually reduces the learning rate over time as the network approaches convergence.

To evaluate the performance of our model, we utilize the standard metrics of recall and specificity. Recall measures the proportion of positive samples that the model correctly identifies, or in other words, the percentage of times within a week from a fault that the model issues a warning. Specificity, on the other hand, reflects the proportion of negative samples that the model accurately predicts, which translates to the frequency with which the model does not wrongly predict a fault when there is none. Specificity has top priority in this task to avoid unnecessary costs from false positives.

Our results, presented in Table I, show that the best recall, namely 0.6694, was achieved with a learning rate of 0.00003 and a learning rate decay of 0.05. On the other hand, the highest specificity, 0.9127, was obtained with a learning rate of 0.0003 and a learning rate decay of 0.05. These findings suggest that the choice of learning rate and learning rate decay can have a significant impact on the performance of the model, and careful tuning of these hyperparameters may be necessary to achieve the best results.

The output of the forecasting model for a short time period is visualized in figure 6, together with two faults and a week-long period before each fault. The model output (blue line) is rising before the faults and dropping down to a lower level after the faults.

## Comparison to baseline

To the best of our knowledge, there have been very limited published attempts to predict faults in the power grid based on disturbance recordings. A related approach is [15], which is using weather data to predict faults. In fact, there are few methods available for predicting discrete events based on time series data in general. One approach related to this is [16], which aims to predict time series *during* extreme events, but this is still significantly different from our task.

Since there are few existing methods for solving our problem, and since our data is unique, there are no baselines for comparison. Therefore, we developed a simple baseline model for predicting faults based on the frequency of recorded anomalies. The model counts the number of recordings during a period of one week, filtered using the first component of our proposed model, and issues a warning if the count exceeds a certain threshold. Using this baseline method and adjusting the threshold to achieve a specificity of 0.9127, we obtained a recall of 0.0889. In comparison, our proposed method achieved a recall of 0.4367 at the same specificity. Adjusting the threshold to instead achieve a recall of 0.6694 resulted in a specificity of 0.3640, which is also significantly lower than that of our proposed method.

Table I. The results from hyperparameter tuning

| Learning rate | Learning rate decay | Test recall | Test specificity |
|---|---|---|---|
| 0.00003 | 0.01 | 0.5316 | 0.8517 |
| 0.00003 | 0.05 | **0.6694** | 0.7495 |
| 0.0001 | 0.01 | 0.4921 | 0.9088 |
| 0.0001 | 0.05 | 0.5550 | 0.8616 |
| 0.0003 | 0.01 | 0.4607 | 0.9102 |
| 0.0003 | 0.05 | 0.4367 | **0.9127** |
| 0.001 | 0.01 | 0.4638 | 0.8955 |
| 0.001 | 0.05 | 0.5111 | 0.8918 |

## Discussion

**Feature engineering**

Our feature engineering serves two important purposes. First, it highlights the parts of the recorded anomalies that are most relevant for the task. Without first extracting key properties from the waveforms, each data point would be very high-dimensional, making it more difficult for the model to learn. In order to use the raw waveforms as input, without any pre-processing, a more complex model would be required, which likely means that more data would be needed for the model to train reliably. While it is possible that this could yield good results, in most practical situations, data is scarce, making data efficiency critical. It is worth noting that overdoing feature engineering can result in the loss of essential information, as seen in the poor performance of the baseline model based on the frequency of recorded anomalies. Therefore, careful development of appropriate feature engineering is crucial.

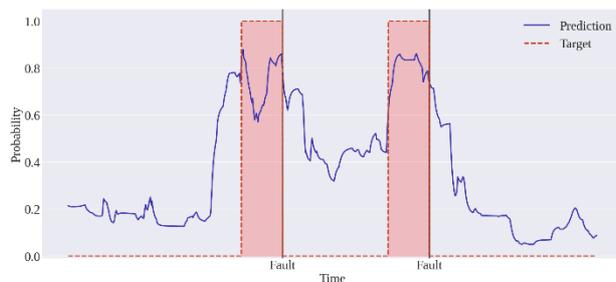

Figure 6: Visualization of a period of time with two faults, along with a one-week period before each fault, as well as the model output for the period. The predicted risk (model output) exhibits a significant increase before each of the two faults occur, enabling an alarm to be raised.

The second reason is that, without significant down sampling of the input data through our feature extraction process, the model would become more computationally expensive, requiring powerful hardware to train. One key advantage of our approach is that it is designed to be efficient enough to run on a standard laptop CPU, something that would not be possible if the model were more complex. This makes it possible to use the model in a wide range of computing environments without needing specialized hardware.

**Importance of filtering**

As with the feature engineering, the filtering of recordings is important in order to not include unnecessary noise when making predictions, but instead focus the attention of the model on the most relevant information. With a more complex model, that is better able to ignore certain information and put extra focus on other parts, the filtering would not be needed. However, for the same reasons as noted previously, this would come at the cost of a more computationally expensive model, which in turn would increase the need for data.

## CONCLUSIONS

In this paper, we proposed a novel deep learning framework for predicting future faults in the power grid. We showed that our LSTM-based model is able to successfully predict faults based only on disturbance recordings. Also, we observed that tuning the parameters and hyperparameters, such as features for creating sequences, or the learning rate of the optimization algorithm, can affect the prediction accuracy. Thus, for an accurate predictor model, four essential steps must be followed: suitable feature selection, feature engineering, innovative selection and structuring of neural networks, and hyperparameter optimization of the machine learning methods.